\documentclass[twocolumn,showpacs,preprintnumbers,amsmath,amssymb]{revtex4}

\usepackage{graphicx}
\usepackage{dcolumn}
\usepackage{bm}

\begin{document}


\title{von Neumann-Landau equation for wave functions,
wave-particle duality and collapses of wave functions}
\author{Zeqian Chen}
\email{zqchen@wipm.ac.cn}
\affiliation{%
State Key Laboratory of Magnetic Resonance and Atomic and Molecular
Physics and United Laboratory of Mathematical Physics, Wuhan
Institute of Physics and Mathematics, Chinese Academy of Sciences,
30 West District, Xiao-Hong-Shan, P.O.Box 71010, Wuhan, China}

\date{\today}

\begin{abstract}
It is shown that von Neumann-Landau equation for wave functions can
present a mathematical formalism of motion of quantum mechanics. The
wave functions of von Neumann-Landau equation for a single particle
are `bipartite', in which the associated Schr\"{o}dinger's wave
functions correspond to those `bipartite' wave functions of product
forms. This formalism establishes a mathematical expression of
wave-particle duality and that von Neumann's entropy is a
quantitative measure of complementarity between wave-like and
particle-like behaviors. Furthermore, this extension of
Schr\"{o}dinger's form suggests that collapses of Schr\"{o}dinger's
wave functions can be regarded as the simultaneous transition of the
particle from many levels to one.
\end{abstract}

\maketitle


\section{Introduction}
As is well known, Heisenberg's equation \cite{H} and
Schr\"{o}dinger's equation \cite{S}, as two forms for the equations
of motion of quantum mechanics, are equivalent. Of these, the
Schr\"{o}dinger form seems to be the more useful one for practical
problems, as it provides differential equations for wave functions,
while Heisenberg's equation involves as unknowns the operators
forming the representative of the dynamical variable, which are far
more numerous and therefore more difficult to evaluate than the
Schr\"{o}dinger unknowns. Also, determining energy levels of various
dynamic systems is an important task in quantum mechanics, for this
solving Schr\"{o}dinger's wave equation is a usual way. Recently,
Fan and Li \cite{FL} showed that Heisenberg's equation can also be
used to deduce the energy level of some systems. By introducing the
conception of invariant `eigen-operator', they derive energy-level
gap formulas for some dynamic Hamiltonians. However, their
`invariant eigen-operator' equation involves operators as unknowns,
as similar to Heisenberg's equation, and hence is also difficult to
evaluate in general.

On the other hand, wave-particle duality, as manifest in the
two-slit experiment, provides perhaps the most vivid illustration of
Bohr's complementarity principle which refers to the ability of
quantum-mechanical entities to behave as waves or particles under
different experiment conditions \cite{Bohr}. Wave-like
(interference) behaviour can be explained by the superposition
principle \cite{Dirac}, while the usual explanation for the loss of
interference (particle-like behaviour) in a which-way experiment is
based on Heisenberg's uncertainty principle \cite{Feynman}. However,
it is demonstrated in the which-way experiment with an atom
interferometer that Heisenberg's position-momentum uncertainty
relation cannot explain the loss of interference for which the
correlations between the which-way detector and the atomic beams are
considered to be responsible \cite{DNR}. From a theoretical point of
view, a measurement-free express of wave-particle duality deserves
research.

In this article, we show that von Neumann-Landau equation for wave
functions (vNLW) is an extension of Schr\"{o}dinger's wave equation
and can be used to determine energy-level gaps of the system.
Contrary to Schr\"{o}dinger's wave equation, vNLW is on `bipartite'
wave functions. It is shown that these `bipartite' wave functions
satisfy all the basic properties of Schr\"{o}dinger's wave functions
which correspond to those `bipartite' wave functions of product
forms. In particular, we will show that this extension of
Schr\"{o}dinger's form establishes a mathematical expression of
wave-particle duality and that von Neumann's entropy is a
quantitative measure of complementarity between wave-like and
particle-like behaviors. Furthermore, from this formalism it is
concluded that collapses of Schr\"{o}dinger's wave functions can be
regarded as the simultaneous transition of the particle from many
levels to one.

\section{von Neumann-Landau equation for wave functions}
Consider the quantum system of a single particle. Note that the
Hamiltonian for a single particle in an external field
is\begin{equation}\label{2.1}\hat{H}(\vec{x}) = - \frac{\hbar^2}{ 2
m} \nabla^2_{\vec{x}} + U(\vec{x} ), \tag{2.1}\end{equation}where
$\nabla^2_{\vec{x}} = \partial^2/\partial x^2_1 +
\partial^2/\partial x^2_2 + \partial^2/\partial x^2_3,$
$U(\vec{x})$ is the potential energy of the particle in the external
field, and $\vec{x} = (x_1, x_2, x_3) \in \mathbf{R}^3.$ The
Schr\"{o}dinger's wave equation describing dynamics of the particle
is\begin{equation}\label{2.2}i\hbar \frac{\partial \psi (\vec{x}, t)
}{\partial t} = \hat{H}(\vec{x}) \psi (\vec{x}, t) = -
\frac{\hbar^2}{ 2 m} \nabla^2_{\vec{x}} \psi (\vec{x}, t) +
U(\vec{x} ) \psi (\vec{x}, t). \tag{2.2}\end{equation}The state of
the particle can be described by a definite wave function $\psi$ of
Equation \eqref{2.2}, whose stationary states determine its energy
levels. Moreover, the expectation value of an observable $\hat{Q}$
in the state corresponding to $\psi$ is determined by $\langle
\hat{Q} \rangle_{\psi} = \langle \psi | \hat{Q} | \psi\rangle.$

On the other hand, let $\psi (\vec{x}, t)$ and $\varphi (\vec{x},
t)$ both satisfy Eq.\eqref{2.2}. Then we
have\begin{equation*}\begin{split} i\hbar &\frac{\partial ( \psi
(\vec{x}, t) \varphi^* (\vec{y}, t)) }{\partial t}\\
& = i\hbar \frac{\partial \psi (\vec{x}, t) }{\partial t} \varphi^*
(\vec{y}, t) + i\hbar \frac{\partial \varphi^* (\vec{y}, t)
}{\partial t} \psi (\vec{x}, t)\\
& = \left [ \hat{H}(\vec{x}) \psi (\vec{x}, t) \right ] \varphi^*
(\vec{y}, t) - \left [ \hat{H}(\vec{y}) \varphi
(\vec{y}, t) \right ]^* \psi (\vec{x}, t)\\
& = \left ( \hat{H}(\vec{x}) - \hat{H}(\vec{y}) \right ) (\psi
(\vec{x}, t) \varphi^* (\vec{y}, t)).
\end{split}\end{equation*}
This leads to the following wave equation
\begin{equation}\label{2.3}i\hbar \frac{\partial \Psi
(\vec{x}, \vec{y}; t) }{\partial t} = \left (\hat{H}(\vec{x}) -
\hat{H}(\vec{y}) \right ) \Psi (\vec{x}, \vec{y}; t),
\tag{2.3}\end{equation}where $\Psi (\vec{x}, \vec{y}; t ) \in
L^2_{\vec{x}, \vec{y}}.$ Contrary to Schr\"{o}dinger's wave equation
Eq.\eqref{2.2} for `one-partite' wave functions $\psi (\vec{x}) \in
L^2_{\vec{x}},$ the wave equation Eq.\eqref{2.3} is an differential
equation for `bipartite' wave functions $\Psi (\vec{x}, \vec{y}),$
which, replacing $\hat{H}(\vec{x}) + \hat{H}(\vec{y})$ by
$\hat{H}(\vec{x}) - \hat{H}(\vec{y}),$ is also different from
Schr\"{o}dinger's wave equation for two particles.

We would like to mention that Eq.\eqref{2.3} has been presented by
von Neumann \cite{vN} and Landau \cite{LL} giving the change in the
density matrix with time. Here, we regard Eq.\eqref{2.3} as a wave
equation but not a equation for density functions. This is the key
point which is distinct from \cite{vN} and \cite{LL}. As follows, we
will show that von Neumann-Landau equation for wave functions
Eq.\eqref{2.3} is an extension of Schr\"{o}dinger's wave equation
Eq.\eqref{2.2} and a suitable form for wave-particle duality.

Since$$\frac{\partial \left | \Psi (\vec{x}, \vec{y}; t) \right
|^2}{\partial t} = 2 \mathrm{Re} \left [ \Psi^* (\vec{x}, \vec{y};
t) \frac{\partial \Psi (\vec{x}, \vec{y}; t) }{\partial t} \right
],$$it is concluded from Eq.\eqref{2.3}
that\begin{equation}\label{2.4} \frac{\partial }{\partial t} \int
\left | \Psi (\vec{x}, \vec{y}; t) \right |^2 d^3\vec{x} d^3 \vec{y}
= 0. \tag{2.4}\end{equation}This implies that Eq.\eqref{2.3}
preserves the density $\left | \Psi (\vec{x}, \vec{y}; t) \right |^2
d^3\vec{x} d^3 \vec{y}$ with respect to time and means that, if this
wave function $\Psi$ is given at some instant, its behavior at all
subsequent instants is determined.

By Schmidt's decomposition theorem \cite{Schmidt}, for every $\Psi
(\vec{x}, \vec{y}) \in L^2_{\vec{x}, \vec{y}}$ there exist two
orthogonal sets $\{\psi_n \}$ and $\{\varphi_n \}$ in
$L^2_{\vec{x}}$ and $L^2_{\vec{y}}$ respectively, and a sequence of
positive numbers $\{ \mu_n \}$ satisfying $\sum_n \mu^2_n  < \infty$
so that\begin{equation}\label{2.5}\Psi (\vec{x}, \vec{y}) = \sum_n
\mu_n \psi_n (\vec{x}) \varphi^*_n (\vec{y}).
\tag{2.5}\end{equation}Then, it is easy to check that$$\Psi
(\vec{x}, \vec{y}; t) = \sum_n \mu_n \psi_n (\vec{x}, t) \varphi^*_n
(\vec{y}, t)$$satisfies Eq.\eqref{2.3} with $\Psi (\vec{x}, \vec{y};
0) = \Psi (\vec{x}, \vec{y}),$ where both $\psi_n (\vec{x}, t)$ and
$\varphi_n (\vec{y}, t)$ satisfy Eq.\eqref{2.2} with $\psi_n
(\vec{x}, 0) = \psi_n (\vec{x})$ and $\varphi_n (\vec{y}, 0) =
\varphi_n (\vec{y}),$ respectively. Hence, the wave equation
Eq.\eqref{2.3} can be solved mathematically from Schr\"{o}dinger's
wave equation.

Also, given $\psi \in L^2_{\vec{x}},$ for every $t \geq 0$ define
operators $\varrho_t$ on $L^2_{\vec{x}}$
by\begin{equation}\label{2.6}(\varrho_t \varphi ) (\vec{x}) = \int
\Psi (\vec{x}, \vec{y}; t) \varphi (\vec{y}) d^3 \vec{y},
\tag{2.6}\end{equation}where $\Psi (\vec{x}, \vec{y}; t)$ is the
solution of Eq.\eqref{2.3} with $\Psi (\vec{x}, \vec{y}; 0) = \psi
(\vec{x}) \psi^* (\vec{y}).$ It is easy to check
that\begin{equation}\label{2.7}i \frac{\partial \varrho_t}{\partial
t } = \left [ H, \varrho_t \right ],~~\varrho_0 = | \psi \rangle
\langle \psi |. \tag{2.7}\end{equation}This is just
Schr\"{o}dinger's equation in the form of density operators. Hence,
Schr\"{o}dinger's wave equation is a special case of the wave
equation Eq.\eqref{2.3} with initial values of product form $\Psi
(\vec{x}, \vec{y}; 0) = \psi (\vec{x}) \psi^* (\vec{y}).$ This
concludes that the wave equation Eq.\eqref{2.3} is an extension of
Schr\"{o}dinger's wave equation.

In the sequel, we consider the problem of stationary states. Let
$\psi_n$ be the eigenfuncions of the Hamiltonian operator $\hat{H},$
i.e., which satisfy the equation\begin{equation}\label{2.8}
\hat{H}(\vec{x}) \psi_n (\vec{x}) = E_n \psi_n (\vec{x}),
\tag{2.8}\end{equation}where $E_n$ are the eigenvalues of $\hat{H}.$
Correspondingly, the wave equation
Eq.\eqref{2.3}\begin{equation*}\begin{split}i\hbar \frac{\partial
\Psi (\vec{x}, \vec{y}; t) }{\partial t} & = \left (\hat{H}(\vec{x})
- \hat{H}(\vec{y}) \right ) \Psi (\vec{x}, \vec{y}; t)\\
& = (E_n - E_m ) \Psi (\vec{x}, \vec{y};
t)\end{split}\end{equation*}with $\Psi (\vec{x}, \vec{y}; 0) =
\psi_n (\vec{x}) \psi^*_m (\vec{y}),$ can be integrated at once with
respect to time and gives\begin{equation}\label{2.9}\Psi (\vec{x},
\vec{y}; t) = e^{-i\frac{1}{\hbar} (E_n - E_m) t} \psi_n (\vec{x})
\psi^*_m (\vec{y}). \tag{2.9}\end{equation}Since $\{ \psi_n
(\vec{x}) \}$ is a complete orthogonal set in $L^2_{\vec{x}},$ it is
concluded that $\{ \psi_n (\vec{x}) \psi^*_m (\vec{y}) \}$ is a
complete orthogonal set in $L^2_{\vec{x}, \vec{y}}.$ Then, for every
$\Psi (\vec{x}, \vec{y}) \in L^2_{\vec{x}, \vec{y}}$ there exists a
unique set of numbers $\{ c_{n,m}\}$ satisfying $\sum_{n,m} |
c_{n,m} |^2 < \infty$ so that
\begin{equation}\label{2.10}\Psi (\vec{x}, \vec{y}) = \sum_{n,m} c_{n,m} \psi_n (\vec{x})
\psi^*_m (\vec{y}). \tag{2.10}\end{equation}Hence, for $\Psi
(\vec{x}, \vec{y}; 0) = \sum_{n,m} c_{n,m} \psi_n (\vec{x}) \psi^*_m
(\vec{y})$ we have that\begin{equation}\label{2.11}\Psi (\vec{x},
\vec{y}; t) = \sum_{n,m} c_{n,m} e^{-i\frac{1}{\hbar} (E_n - E_m) t}
\psi_n (\vec{x}) \psi^*_m (\vec{y}) \tag{2.11}\end{equation}for $t
\geq 0.$

Conversely, if $\Psi (\vec{x}, \vec{y}) \in L^2_{\vec{x}, \vec{y}}$
is an eigenfuncion of the operator $\hat{H}(\vec{x})-
\hat{H}(\vec{y}),$ i.e., which satisfies the
equation\begin{equation}\label{2.12} \left ( \hat{H}(\vec{x})-
\hat{H}(\vec{y}) \right ) \Psi (\vec{x}, \vec{y}) = \lambda \Psi
(\vec{x}, \vec{y}), \tag{2.12}\end{equation}where $\lambda$ is an
associated eigenvalue, then $\Psi (\vec{x}, \vec{y}; t) =
e^{-i\frac{1}{\hbar} \lambda t} \Psi (\vec{x}, \vec{y})$ satisfies
Eq.\eqref{2.3} and consequently, it is concluded from
Eq.\eqref{2.11} that $\lambda = E_n - E_m$ is an energy-level gap of
the system. Thus, the wave equation Eq.\eqref{2.3} can be used to
determine energy-level gaps of the system Eq.\eqref{2.2}.

\section{quantum measurement}
It is well known that the basis of the mathematical formalism of
quantum mechanics lies in the proposition that the state of a system
can be described by a definite Schr\"{o}dinger's wave function of
coordinates \cite{vN,Dirac}. As an extension of Schr\"{o}dinger's
wave functions, it seems that the state of a quantum system also can
be described by a definite `bipartite' wave function of
Eq.\eqref{2.3}, of which the physical meaning is that the
`bipartite' wave functions of stationary states determine
energy-level gaps of the system.

In fact, we can make the general assumption that if the measurement
of an observable $\hat{O}$ for the system in the `bipartite' state
corresponding to $\Psi$ is made a large number of times, the average
of all the results obtained will be\begin{equation}\label{3.1}
\langle \hat{O} \rangle_{\Psi} = \mathrm{Tr} \left [ \varrho_{\Psi}
\hat{O} \varrho^{\dagger}_{\Psi} \right ],
\tag{3.1}\end{equation}where $\varrho_{\Psi}$ is an operator on
$L^2$ associated with $\Psi$ defined
by\begin{equation}\label{3.2}(\varrho_{\Psi} \varphi ) (\vec{x}) =
\int \Psi (\vec{x}, \vec{y}) \varphi (\vec{y}) d^3
\vec{y}\tag{3.2}\end{equation}for every $\varphi \in L^2,$ provided
$\Psi$ is normalized since\begin{equation}\label{3.3}\mathrm{Tr
\varrho^{\dagger}_{\Psi} \varrho_{\Psi}} = \int \bigl | \Psi
(\vec{x}, \vec{y}) \bigr |^2 d^3 \vec{x} d^3 \vec{y} =
1.\tag{3.3}\end{equation}That is, the expectation value of an
observable $\hat{O}$ in the `bipartite' state corresponding to
$\Psi$ is determined by Eq.\eqref{3.1}. In particular, the
probability of the system in the `bipartite' state corresponding to
$\Psi$ reduces to a state $| \varphi \rangle$ after measurement
is\begin{equation}\label{3.4}\begin{split} \langle | \varphi \rangle
\langle \varphi| \rangle_{\Psi} & = \mathrm{Tr} \left [
\varrho_{\Psi} | \varphi \rangle \langle \varphi|
\varrho^{\dagger}_{\Psi} \right ]\\
& = \langle \varphi, \varrho^{\dagger}_{\Psi} \varrho_{\Psi} \varphi
\rangle\\
& = \| \varrho_{\Psi} \varphi \|^2\\
& = \int \left | \int \Psi (\vec{x}, \vec{y}) \varphi (\vec{y}) d^3
\vec{y} \right |^2 d^3 \vec{x}.\end{split}\tag{3.4}\end{equation} It
is easy to check that if $\Psi (\vec{x}, \vec{y}) = \psi (\vec{x})
\psi^* (\vec{y}),$ then $\varrho_{\Psi} = | \psi \rangle \langle
\psi |$ and so\begin{equation}\label{3.5} \langle \hat{O}
\rangle_{\Psi} = \langle \psi | \hat{O} | \psi \rangle.
\tag{3.5}\end{equation}This concludes that our expression
Eq.\eqref{3.1} agrees with the interpretation of Schr\"{o}dinger's
wave functions for calculating expectation values of any chosen
observable.

\section{Wave-particle duality}
Since every normalized `bipartite' wave function $\Psi (\vec{x},
\vec{y})$ is of the form \eqref{2.5} with $\sum_n \mu^2_n  = 1,$ we
can define the `entanglement' measure of `bipartite' wave functions
$\Psi$ by\begin{equation}\label{4.1} S (\Psi) = - \sum_n \mu^2_n \ln
\mu^2_n. \tag{4.1}\end{equation}It is easy to show
that\begin{equation}\label{4.2} S (\Psi) = - \mathrm{tr} \left [
\varrho_{\vec{x}} (\Psi) \ln \varrho_{\vec{x}} (\Psi) \right ] = -
\mathrm{tr} \left [ \varrho_{\vec{y}} (\Psi) \ln \varrho_{\vec{y}}
(\Psi) \right ] , \tag{4.2} \end{equation}where $\varrho_{\vec{x}}
(\Psi) = \mathrm{tr}_{\vec{y}} \left ( | \Psi \rangle \langle \Psi |
\right )$ and $\varrho_{\vec{y}} (\Psi) = \mathrm{tr}_{\vec{x}}
\left ( | \Psi \rangle \langle \Psi | \right ).$ Hence, $S (\Psi)$
is the von Neumann's entropy of the reduced density matrix
$\varrho_{\vec{x}} (\Psi)$ (or equivalently, $\varrho_{\vec{y}}
(\Psi)$) of $| \Psi \rangle \langle \Psi |$ (e.g., \cite{BDSW}). In
the sequel, we will show that $S (\Psi)$ is a quantitative measure
of complementarity between wave-like and particle-like behaviors.

Let us imagine a screen impermeable to electrons, in which two
slits, 1 and 2, are cut. We denote by $\psi_1$ the wave function of
an electron through slit 1 with slit 2 being covered, and $\psi_2$
the wave function of an electron through slit 2 with slit 1 being
covered. Then, the state of a single electron through slits 1 and 2
can be described by a `bipartite' wave function of
form\begin{equation}\label{4.3}\begin{array}{lcl} \Psi (\vec{x},
\vec{y}) &=& a_{11} \psi_1 ( \vec{x} ) \psi^*_1 (\vec{y}) + a_{12}
\psi_1 ( \vec{x} ) \psi^*_2 (\vec{y})\\[0.4cm]&~&~~ + a_{21} \psi_2 ( \vec{x} ) \psi^*_1
(\vec{y}) + a_{22} \psi_2 ( \vec{x} ) \psi^*_2 (\vec{y}),\end{array}
\tag{4.3}\end{equation}where $|a_{11}|^2 + |a_{12}|^2 + |a_{21}|^2 +
|a_{22}|^2 =1.$ As following are two special cases of
\eqref{4.3}:\begin{equation}\label{4.4}\Psi_W (\vec{x}, \vec{y}) =
\frac{1}{2} \left [  \psi_1 ( \vec{x} ) + \psi_2 ( \vec{x} ) \right
] \left [ \psi^*_1 ( \vec{y} ) + \psi^*_2 ( \vec{y} ) \right
],\tag{4.4}\end{equation}and
\begin{equation}\label{4.5}\Psi_P (\vec{x}, \vec{y}) = \frac{1}{\sqrt{2}} \left
[ \psi_1 ( \vec{x} ) \psi^*_1 ( \vec{y} ) +  \psi_2 ( \vec{x} )
\psi^*_2 ( \vec{y} ) \right ].\tag{4.5}\end{equation}Accordingly,
$\Psi_W$ corresponds to Schr\"{o}dinger's wave function $\psi =
\frac{1}{\sqrt{2}} (\psi_1 + \psi_2),$ while there is no
Schr\"{o}dinger's wave function associated with $\Psi_P.$ A single
electron described by $\Psi_W$ behaves like waves, while by $\Psi_P$
like particles. This is so because for position, by \eqref{3.1} we
have\begin{equation}\label{4.6}\langle \hat{x} \rangle_{\Psi_W} =
\frac{1}{2} \left | \psi_1 ( \vec{x}) + \psi_2 ( \vec{x}) \right |^2
,\tag{4.6}\end{equation} and\begin{equation}\label{4.7} \langle
\hat{x} \rangle_{\Psi_P} = \frac{1}{2} \left ( | \psi_1 (
\vec{x})|^2 + | \psi_2 ( \vec{x})|^2 \right
),\tag{4.7}\end{equation}respectively. Generally, for every $\Psi$
of \eqref{4.3} one has$$0 \leq S ( \Psi ) \leq S ( \Psi_P ) =
\frac{1}{2} \ln 2.$$When a single electron is described by $\Psi,$
the larger is $S ( \Psi ),$ more like particles it behaves. Hence,
$S ( \Psi )$ characterizes quantitatively wave-particle duality for
a single particle. Moreover, $S ( \Psi )$ characterizes
quantitatively the path entanglement of a single particle in the
which-way experiment \cite{BSHMU}.

\section{Collapses of wave functions}
Let $\{ \psi_n \}$ be the eigenfuncions of the Hamiltonian operator
$\hat{H}.$ Then, every normalized `bipartite' wave function $\Psi
(\vec{x}, \vec{y})$ can be expressed as \eqref{2.10} with
$\sum_{n,m} | c_{n,m} |^2 = 1.$ Note that$$ \left (
\hat{H}(\vec{x})- \hat{H}(\vec{y}) \right ) \left [ \psi_n (\vec{x})
\psi^*_m (\vec{y}) \right ] = (E_n - E_m) \left [ \psi_n (\vec{x})
\psi^*_m (\vec{y}) \right ].$$Then, $c_{n,m}$ can be regarded as the
probabilistic amplitude of the transition of the particle from level
$\psi_n$ to $\psi_m.$ The probability of getting $E_m$ on
measurement in a state with `bipartite' wave function $\Psi$
is\begin{equation}\label{5.1}p_m = \sum_n | c_{n,m}|^2.
\tag{5.1}\end{equation}The value of the associated change of energy
of the system is\begin{equation}\label{5.2}\triangle E_m = \sum_n |
c_{n,m}|^2 ( E_n - E_m ).\tag{5.2}\end{equation}In particular, if
$\Psi (\vec{x}, \vec{y}) = \psi (\vec{x}) \psi^* (\vec{y})$ with
$\psi = \sum_n a_n \psi_n$ and $\sum_n |a_n|^2 = 1,$ then $c_{n,m} =
a_n a^*_m$ and hence $p_m = |a_m|^2.$ In this case, the collapse of
$\psi$ to $\psi_m$ can be regarded as the simultaneous transition of
the particle from levels $\psi_1, \psi_2, \ldots$ to $\psi_m.$ Thus,
our results suggest that von Neumann's collapse of Schr\"{o}dinger's
wave functions is just the simultaneous transition of the particle
from many levels to one \cite{vN}. We therefore conclude that there
are two basic changes for the system of a single particle, one is
unitary change, while the other is von Neumann's collapse of wave
functions in such a sense that it is the simultaneous transition of
the particle from many levels (perhaps, only one) to one.

\section{Conclusion}
In conclusion, we show that von Neumann-Landau equation for wave
functions is an extension of Schr\"{o}dinger's wave equation and can
be used to determine energy-level gaps of the system for a single
particle. It is presented a mathematical expression of wave-particle
duality and that von Neumann's entropy is a quantitative measure of
complementarity between wave-like and particle-like behaviors.
Moreover, our formalism suggests that von Neumann's collapse of
Schr\"{o}dinger's wave functions is just the simultaneous transition
of the particle from many levels to one.

I am grateful to Prof.Dr.Albeverio for inviting me to visit
Institute for Applied Mathematics, Bonn University, in January,
2007, where some results of the paper were reported. This work was
partially supported by the National Natural Science Foundation of
China under Grant No.10571176 and No.10775175.



\begin{thebibliography}{**}
\bibitem{H}W.Heisenberg, Z.Physik {\bf 33}, 879(1925).
\bibitem{S}E.Schr\"{o}dinger, Ann.Physik {\bf 79}, 36(1926).
\bibitem{FL}H.-Y.Fan and C.Li, Phys.Lett.A {\bf 321}, 75(2004).
\bibitem{Bohr}N.Bohr, {\it Albert Einstein: Philosopher-Scientist} (ed. Schilpp, P.A.)
200-241 (Library of Living Philosophers, Evanston, 1949).
\bibitem{Dirac}P.A.M.Dirac, {\it The Principles of Quantum
Mechanics} (Fourth edition, Oxford University Press, Oxford, 1958).
\bibitem{Feynman}R.P.Feynman and A.R.Hibbs, {\it Quantum Mechanics and
Path Integrals} (McGraw-Hill, Inc., 1965).
\bibitem{DNR}S.D\"{u}rr, T.Nonn, and G.Rempe Nature
{\bf 395}, 33(1998).
\bibitem{vN}J.von Neumann, {\it Mathematical Foundations of Quantum Mechanics}
(Princeton University Press, Princeton, 1955).
\bibitem{LL}L.D.Landau and E.M.Lifshitz, {\it Quantum Mechanics
$($Non-relativistic Theory$)$} (Third edition, Pergamon Press,
Oxford, 1987).
\bibitem{Schmidt}E.Schmidt, Math.Ann. {\bf 63}, 433(1907).
\bibitem{BDSW}C.H.Bennett, D.P.DiVincenzo, J.Smolin, and
W.K.Wootters, Phys.Rev.A {\bf 54}, 3814(1996).
\bibitem{BSHMU}E.Buks, S.Schuster, M.Heiblum, D.Mahalu, and
V.Umansky, Nature {\bf 391}, 871(1998).
\end{thebibliography}
\end{document}